\documentclass[prb,twocolumn,a4paper,showpacs,floatfix]{revtex4} 
\usepackage{latexsym}
\usepackage{graphicx,rotating}
\newif\ifshowfigures \showfigurestrue
\begin{document}

\title{Renormalization group approach to energy level statistics\\ 
  at the integer quantum Hall transition}

\author{Philipp Cain and Rudolf A.\ R\"omer} 

\altaffiliation%
[Permanent address: ]
{Department of Physics, University of Warwick, Coventry CV4
  7AL, UK, Email: r.roemer@warwick.ac.uk} \affiliation{Institut f\"ur
  Physik, Technische Universit\"at Chemnitz, D-09107 Chemnitz,
  Germany}
\author{Mikhail E.\ Raikh} \affiliation{Department of Physics,
  University of Utah, Salt Lake City, Utah 84112}

\date{$Revision: 1.24 $, compiled \today}

\begin{abstract}
  We extend the real-space renormalization group (RG) approach to the
  study of the energy level statistics at the integer quantum Hall
  (QH) transition. Previously it was demonstrated that the RG approach
  reproduces the critical distribution of the {\em power} transmission
  coefficients, i.e., two-terminal conductances, $P_{\text c}(G)$, with very
  high accuracy. The RG flow of $P(G)$ at energies away from the
  transition yielded the value of the critical exponent, $\nu$, that
  agreed with most accurate large-size lattice simulations.  To obtain
  the information about the level statistics from the RG approach, we
  analyze the evolution of the distribution of {\em phases} of the
  {\em amplitude} transmission coefficient upon a step of the RG
  transformation. From the fixed point of this transformation we
  extract the critical level spacing distribution (LSD). This
  distribution is close, but distinctively different from the earlier
  large-scale simulations.  We find that away from the transition the
  LSD crosses over towards the Poisson distribution.  Studying the
  change of the LSD around the QH transition, we check that it indeed
  obeys scaling behavior.  This enables us to use the alternative
  approach to extracting the critical exponent, based on the LSD, and
  to find $\nu=2.37\pm0.02$ very close to the value established in the
  literature. This provides additional evidence for the surprising
  fact that a small RG unit, containing only five nodes, accurately
  captures most of the correlations responsible for the
  localization-delocalization transition.
\end{abstract}


\maketitle

\section{Introduction}

It has been realized long ago that, alongside with the change in the
behavior of the eigenfunctions, a localization-delocalization
transition manifests itself in the statistics of the energy levels. In
particular, as the energy is swept across the mobility edge, the shape
of the level spacing distribution (LSD) crosses over from the
Wigner-Dyson distribution, corresponding to the appropriate
universality class, to the Poisson distribution.

Moreover, finite-size corrections to the critical LSD exactly at the
mobility edge allow to determine the value of the correlation length
exponent,\cite{ShkSSL93} thus avoiding an actual analysis of the
spatial extent of the wave functions.  For this reason, the energy
level statistics constitutes an alternative to the
MacKinnon-Kramer\cite{PicS81a,PicS81b,MacK81,MacK83} and to the
transmission-matrix\cite{Lan70,FisL81} approaches to the numerical
study of localization.

Another reason why a large number of numerical simulations of the LSD
at the transition
\cite{HofS94b,Eva94,VarHSP95,KawOSO96,BatSZK96,ZhaK97,Met98b,Eva95,SchZ95,FeiAB95,OhtO95,BatS96,MetV98,BatSK98,Met99}
were carried out during the past decade is the controversy that
existed over the large-spacing tail of the critical LSD.  Conclusive
demonstration\cite{BatSZK96,ZhaK97,BatSK98} that this tail is
Poissonian, i.e., that there is no repulsion between the levels with
spacings much larger than the mean value,\cite{ShkSSL93} rather than
super-Poissonian,\cite{KraLAA94} implying that repulsion is partially
preserved, required a very high accuracy of the
simulations.\cite{ZhaK97,BatSK98} The bulk of numerical work on the
level statistics at the transition was carried out for
three-dimensional systems
\cite{HofS94b,Eva94,VarHSP95,KawOSO96,BatSZK96,ZhaK97,Met98b} for
which there exists a mobility edge separating localized and extended
states.  In two dimensions all the states are localized in the absence
of a magnetic field. In the presence of a magnetic field,
localization-delocalization transitions in two dimensions (quantum
Hall transitions) are infinitely sharp. Still the reasoning of Ref.\ 
\onlinecite{ShkSSL93} applies.  Numerical studies have established a
Poissonian tail of the LSD.\cite{BatS96} It was also demonstrated
\cite{BatS96} that the procedure of extracting the localization length
exponent from the finite-size corrections yields a value close to $\nu
= 2.35$ found from large-size simulations of the wave
functions.\cite{HucK90,Huc92,Huc95,LeeWK93}

Recently, a semianalytical description of the integer quantum Hall
transition, based on the extension of the scaling ideas for the
classical percolation \cite{StaA92} to the Chalker-Coddington (CC)
model of the quantum percolation,\cite{ChaC88} has been
developed.\cite{GalR97,AroJS97} The key idea of this description, a
real-space-renormalization group approach (RG), is the following.
Each RG step corresponds to a doubling of the system size. The RG
transformation relates the conductance {\em distribution} of the
sample at the next step to the conductance distribution at the
previous step. The {\em fixed point} of this transformation, yields
the distribution of the conductance, $P_{\text c}(G)$ of a {\em
  macroscopic} sample at the quantum Hall transition. This {\em
  universal} distribution describes the mesoscopic properties of a
fully coherent quantum Hall sample.  Analogously to the classical
percolation,\cite{StaA92} the correlation length exponent, $\nu$, was
extracted from the RG procedure \cite{CaiRSR01} using the fact that a
slight shift of the initial distribution with respect to the
fixed-point, $P_{\text c}(G)$, drives the system to the insulator upon
renormalization. Then the rate of the shift of the distribution
maximum determines the value of $\nu$.  Remarkably, both $P_{\text
  c}(G)$ and the critical exponent obtained within the RG
approach\cite{CaiRSR01,WeyJ98,JanMMW98} agree very well with the
``exact'' results of the large-scale
simulations.\cite{Huc92,LeeWK93,WanJL96,WanLS98,AviBB99}
 
The goal of the present paper is two-fold. Firstly, we extend the RG
approach to the level statistics at the transition in order to subject
its validity to yet another test.  Secondly, we apply the method
analogous to the finite-size-corrections analysis to extract $\nu$
from the LSD obtained within the RG approach. This method yields $\nu
= 2.37 \pm 0.02$, which is even closer to the most precise large-scale
simulations result $\nu = 2.35 \pm 0.03$ \cite{Huc92} than the value
$\nu = 2.39 \pm 0.01$ inferred from the conductance
distribution.\cite{CaiRSR01} The latter result is by no means trivial.
Indeed, the original RG transformation \cite{CaiRSR01} related the
conductances, i.e., the {\em absolute values} of the
transmission coefficients of the original and the doubled samples,
while the {\em phases} of the transmission coefficients were assumed
random and uncorrelated.  In contrast, the level statistics at the
transition corresponds to the fixed point in the distribution of these
phases. Therefore, the success of the RG approach for conductances
does not guarantee that it will be equally accurate {\em
  quantitatively} for the level statistics.

Within both RG transformations, for the magnitudes and for the phases
of the transmission coefficients, an initial deviation from the
critical distribution drives the system towards an insulator with zero
transmission and Poissonian LSD.  Thus the procedures of the
extraction of $\nu$ from both transformations are technically
different, but conceptually similar.  In fact, the shape of the
critical LSD, obtained from the RG approach, shows systematic
deviations from the large-scale simulation
results\cite{FeiAB95,OhtO95,BatS96,MetV98,BatSK98,Met99,OnoOK96,KleM97}
which yield the body of LSD very close to the Gaussian unitary random
matrix ensemble (GUE).\cite{Meh90} However, the RG flow of the LSD
towards the insulator appears to be robust.

The paper is organized as follows.  First, in Sec.\ \ref{sec-RG} we
review the real-space RG approach \cite{GalR97,AroJS97,CaiRSR01} and
adjust it to the computation of the energy levels and the LSD.  In
Sec.\ \ref{sec-num} we present our numerical results for the LSD.  The
finite-size scaling (FSS) analysis of the obtained LSD at the QH
transition is reported in Sec.\ \ref{sec-FSS}. Concluding remarks are
presented in Sec.\ \ref{sec-sum}.

\begin{figure}[t]
\ifshowfigures%
\centerline{\includegraphics[width=0.95\columnwidth]{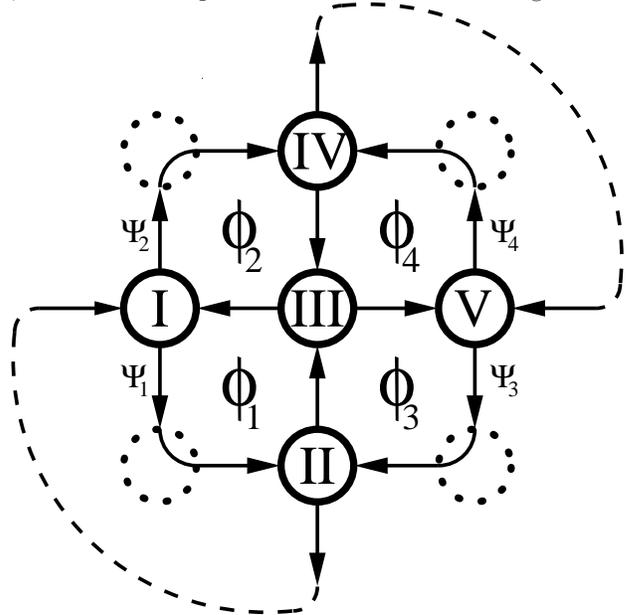}}
\else\centerline{\fbox{\parbox{\columnwidth}{\rule{0cm}{20ex}}}}\fi
\caption{\label{fig-RGstruct}
  Chalker-Coddington network on a square lattice consisting of nodes
  (circles) and links (arrows). The RG unit used for Eq.\ 
  (\ref{eq-qhrg}) combines five nodes (full circles) by
  neglecting some connectivity (dashed circles).  $\Phi_1, \ldots,
  \Phi_4$ are the phases acquired by an electron  along the
  loops as indicated by the arrows. $\Psi_1, \ldots, \Psi_4$ represent
  wave function amplitudes, and the thin dashed lines illustrate the
  boundary conditions used for the computation of level statistics.}
\end{figure}
\section{Model and RG method for the LSD}
\label{sec-RG}

\subsection{RG approach to the conductance distribution}
\label{ssec-RG-RG}

A detailed description of the RG approach to the conductance
distribution can be found in Refs.\ 
\onlinecite{GalR97,AroJS97,CaiRSR01}.  It is based on the RG unit
shown in Fig.\ \ref{fig-RGstruct}.  The unit is a fragment of the CC
network consisting of five nodes.  Each node, $i$, is characterized by
the transmission coefficient $t_i$, which is an amplitude to deflect
an incoming electron along the link to the left. Analogously, the
reflection coefficient $r_i = (1-t_i^2)^{1/2}$ is the amplitude to
deflect the incoming electron to the right.  Doubling of the sample
size corresponds to the replacement of the RG unit by a single node.
The RG transformation expresses the transmission coefficient of this
effective node, $t^{\prime}$, through the transmission coefficients of
the five constituting nodes\cite{GalR97}
\begin{widetext}
\begin{equation}
  \label{eq-qhrg}
  t'= \left | \frac{ t_1 t_5 (r_2 r_3 r_4 e^{i \Phi_2} - 1) + t_2
      t_4 e^{i  (\Phi_3+\Phi_4)} (r_1 r_3 r_5 e^{-i \Phi_1} -
      1) + t_3 (t_2 t_5 e^{i \Phi_3} + t_1 t_4 e^{i \Phi_4}) }
    { (r_3 - r_2 r_4 e^{i \Phi_2}) (r_3 - r_1 r_5 e^{i 
        \Phi_1}) + (t_3 - t_4 t_5 e^{i \Phi_4}) (t_3 - t_1 t_2
      e^{i \Phi_3}) }\right | \quad .
\end{equation}
\end{widetext}
Here $\Phi_j$ are the phases accumulated along the closed loops (see
Fig.\ \ref{fig-RGstruct}). Within the RG approach to the conductance
distribution, information about electron energy is incorporated only
into the values of $t_i$. The energy dependence of phases, $\Phi_j$,
is irrelevant; they are assumed completely random.  Due to this
randomness, the transmission coefficients, $t_i$, for a given energy,
are also randomly distributed with a distribution function $P(t)$.
Then the transformation (\ref{eq-qhrg}) allows, upon averaging
over $\Phi_j$, to generate the next-step distribution $P(t^{\prime})$.
Therefore, within the RG scheme, a delocalized state corresponds to
the fixed point, $P_{\text c}(t)$, of the RG transformation. Due to
the symmetry of the RG unit, it is obvious that the critical
distribution, $P_{\rm c}(t^2)$, of the power transmission coefficient,
$t^2=G$, which has the meaning of the two-terminal conductance, is
symmetric with respect to $t^2=\frac{1}{{2}}$. In other words, the RG
transformation respects the duality between transmission and
reflection.  The critical distribution $P_{\rm c}(G)$ found in Refs.\ 
\onlinecite{GalR97} and \onlinecite{CaiRSR01} agrees very well with
the results of direct large-scale simulations.

\subsection{RG approach to the LSD}

Universal features of the energy level statistics in a macroscopic
fully coherent sample at the quantum Hall transition complement the
universality in the conductance distribution. The prime
characteristics of the level statistics is the LSD -- the distribution
of the spacings between neighboring energy levels. In order to
adjust the RG approach to the calculation of the LSD, it is necessary
to ``close'' the sample at each RG step in order to discretize the
energy levels. One of the possible variants of such a closing is shown
in Fig.\ \ref{fig-RGstruct} with dashed lines.
   
For a given closed RG unit with a fixed set of $t_i$-values at the
nodes, the positions of the energy levels are determined by the energy
dependences, $\Phi_j(E)$, of the four phases along the loops. These
phases change by $\sim \pi$ within a very narrow energy interval,
inversely proportional to the sample size.  Within this interval the
change of the transmission coefficients is negligibly small. A closed
RG unit in Fig.\ \ref{fig-RGstruct} contains $10$ links, and, thus, it
is described by $10$ amplitudes. These amplitudes are related by $10$
equations ($2$ at each node). Each link is characterized by an
individual phase. On the other hand, it is obvious that the energy
levels are determined only by the phases along the loops. One possible
way to derive the system, in which individual phases combine into
$\Phi_j$ is to exclude from the original system of $10$ equations all
amplitudes except the ``boundary'' amplitudes $\Psi_j$ (see Fig.\
\ref{fig-RGstruct}).  This procedure is similar to the derivation of
Eq.\ (\ref {eq-qhrg}). The system of equations for the remaining four
amplitudes takes the form
\begin{widetext}
\begin{equation}\label{system}
\left(
\begin{array}{cccc}
(r_1 r_2 - t_1 t_2 t_3) e^{-i \Phi_1}&
(t_1 r_2 + t_2 t_3 r_1) e^{-i \Phi_1}&
t_2 t_5 r_3 e^{-i \Phi_1}&
t_2 r_3 r_5 e^{-i \Phi_1}\\
-t_1 r_3 r_4 e^{-i \Phi_2}&
r_1 r_3 r_4 e^{-i \Phi_2}&
-(t_4 r_5 + t_3 t_5 r_4) e^{-i \Phi_2}&
(t_4 t_5 - t_3 r_4 r_5) e^{-i \Phi_2}\\
-t_1 t_4 r_3 e^{-i \Phi_4}&
t_4 r_1 r_3 e^{-i \Phi_4}&
(r_4 r_5 - t_3 t_4 t_5) e^{-i \Phi_4}&
-(t_5 r_4 + t_3 t_4 r_5) e^{-i \Phi_4}\\
-(t_2 r_1 + t_1 t_3 r_2) e^{-i \Phi_3}&
-(t_1 t_2 - t_3 r_1 r_2) e^{-i \Phi_3}&
t_5 r_2 r_3 e^{-i \Phi_3}&
r_2 r_3 r_5 e^{-i \Phi_3}
\end{array}
\right)
\left(
\begin{array}{c}
\Psi_1\\
\Psi_2\\
\Psi_3\\
\Psi_4\\
\end{array}
\right)
=
e^{i \omega}
\left(
\begin{array}{c}
\Psi_1\\
\Psi_2\\
\Psi_3\\
\Psi_4\\
\end{array}
\right),
\end{equation}
\end{widetext}
where the parameter $\omega$ should be set zero. Then the energy
levels, $E_k$, of the closed RG unit are the energies for which, with
phases $\Phi_j(E)=\Phi_j(E_k)$, one of the four eigenvalues of the
matrix in the left-hand side of Eq.\ (\ref{system}) is equal to one. If
we keep $\omega$ in the right-hand side of Eq.\ (\ref{system}), then
the above condition can be reformulated as $\omega(E_k)=0$.  Thus, the
calculation of the energy levels reduces to a diagonalization of the
$4\times 4$ matrix.

The crucial step now is the choice of the dependence $\Phi_j(E)$.
If each loop in Fig.\ \ref{fig-RGstruct} is viewed as a closed equipotential
as it is the case for the first step of the RG procedure,\cite{ChaC88} 
then
$\Phi_j(E)$ is a true magnetic phase, which changes linearly
with energy with a slope governed by the actual potential 
profile, which, in turn, determines the drift velocity. Thus we
have  
\begin{equation}
\label{eq-PhiE}
\Phi_j(E)=\Phi_{0,j}+2\pi\frac{E}{s_j},
\end{equation}
where a random part, $\Phi_{0,j}$, is uniformly distributed within
$[0, 2\pi]$, and $2\pi/s_j$ is a random slope. Strictly speaking, the
dependence (\ref{eq-PhiE}) applies only for the first RG step.
At each following step, $n>1$, $\Phi_j(E)$ is a complicated function
of $E$ which carries information about all energy scales at previous
steps. However, in the spirit of the RG approach, we assume that
$\Phi_j(E)$ can still be linearized within a relevant energy interval.
The conventional RG approach suggests that different scales in the
{\em real} space can be decoupled.  Linearization of Eq.\ 
(\ref{eq-PhiE}) implies a similar decoupling in the {\em energy}
space.  In the case of phases, a ``justification'' of such a
decoupling is that at each following RG step, the relevant energy
scale, that is the mean level spacing, reduces by a factor of $4$.

With $\Phi_j(E)$ given by Eq.\ (\ref{eq-PhiE}) and fixed values of
$t_i$, the statistics of energy levels determined by the matrix
equation (\ref{system}) is obtained by averaging over the random
initial phases $\Phi_{0,j}$. In particular, each realization of
$\Phi_{0,j}$ yields 3 level spacings which are then used to construct
a smooth LSD.  We now outline the RG procedure for the LSD. The slopes
$s_j$ in Eq.\ (\ref{eq-PhiE}) determine the level spacings at the
first step. They are randomly distributed with a distribution function
$P_0(s)$. Diagonalization of the matrix in Eq.\ (\ref{system}) with
subsequent averaging over realizations yields the LSD, $P_1(s)$, at
the second step. Then the key element of the RG procedure, as applied
to the level statistics, is using $P_1(s)$ as a {\em distribution of
  slopes} in Eq.\ (\ref{eq-PhiE}).  This leads to the next-step LSD
and so on.

\begin{figure}
\ifshowfigures%
\centerline{\includegraphics[width=0.95\columnwidth]{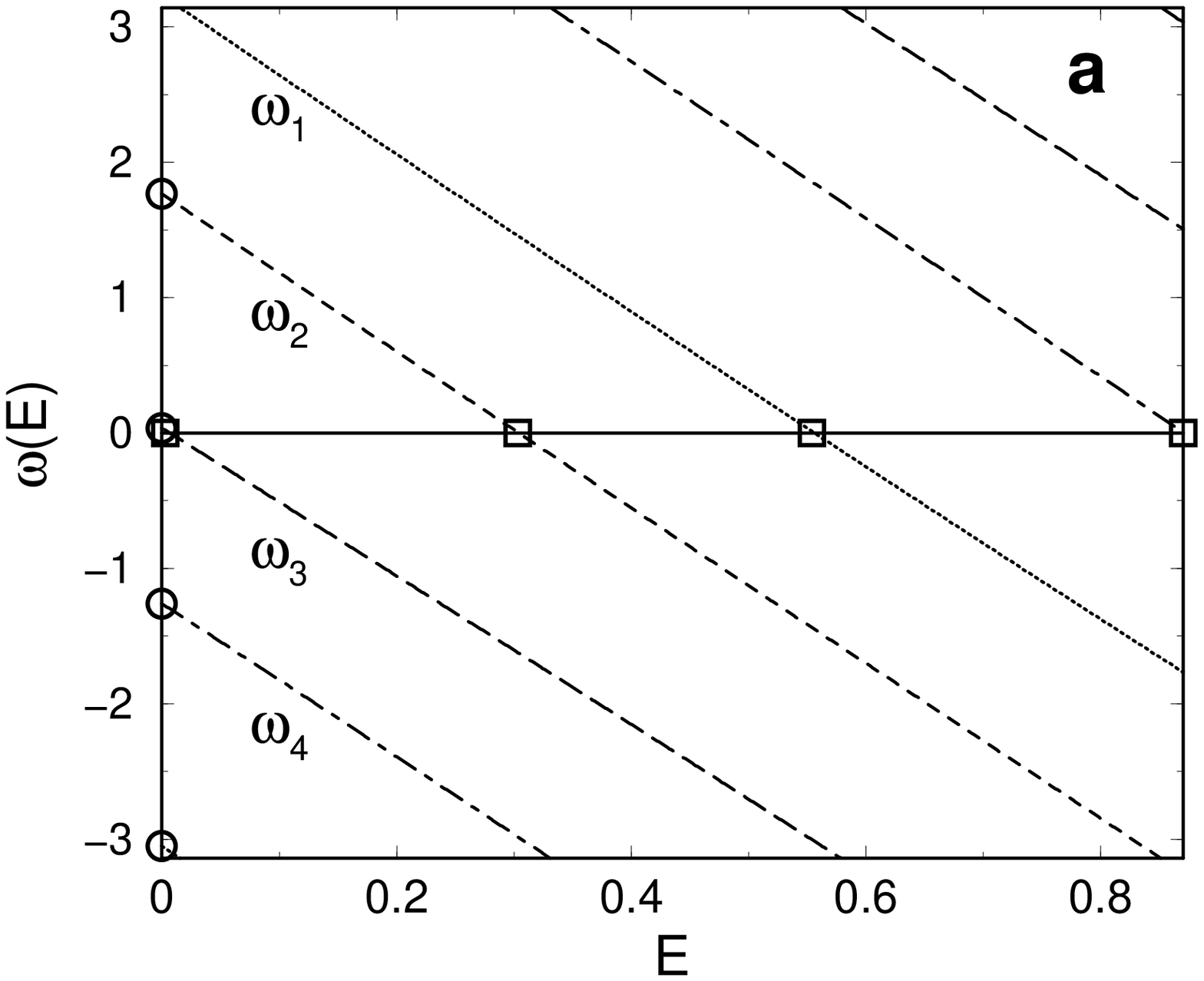}}
\centerline{\includegraphics[width=0.95\columnwidth]{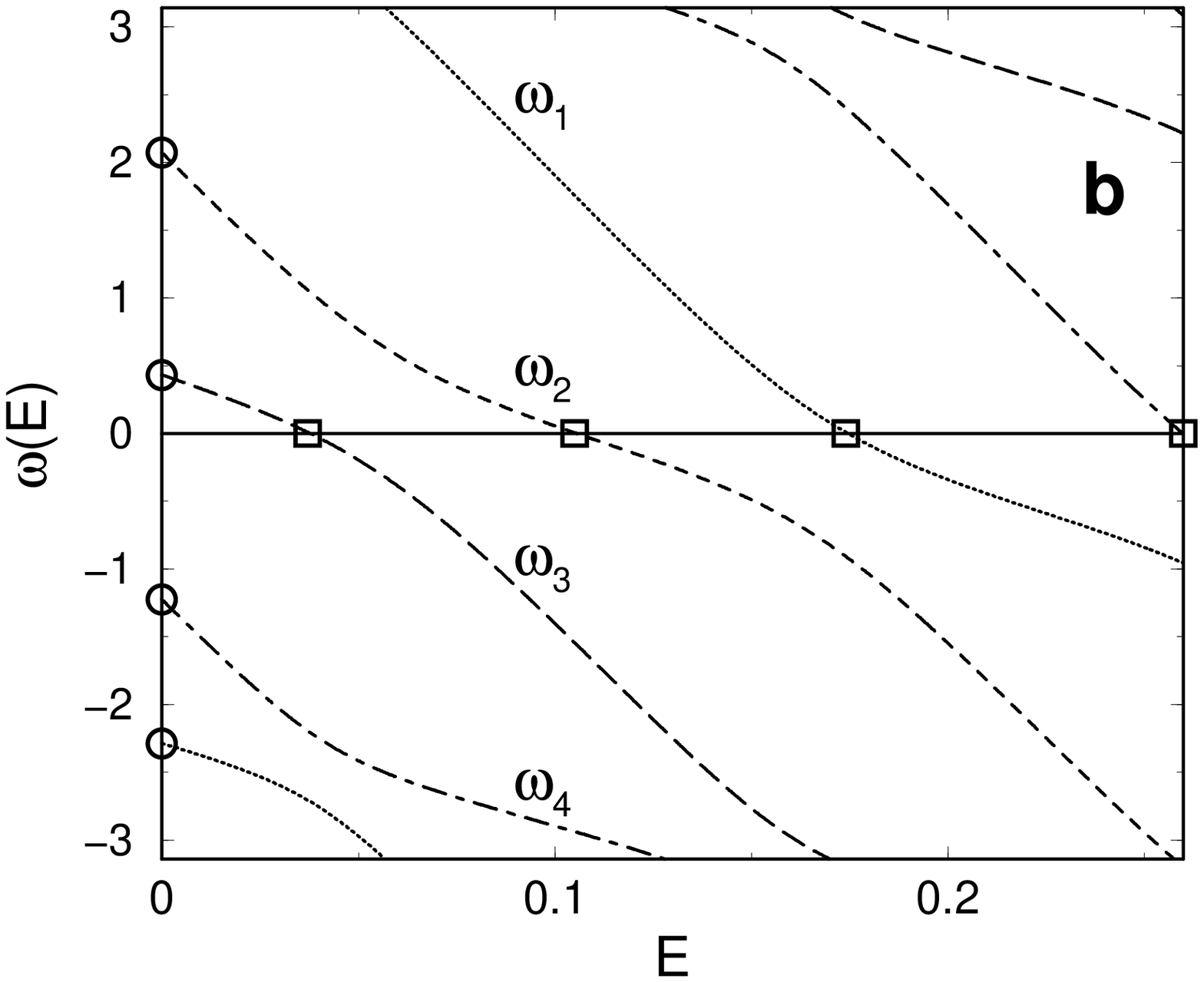}}
\else\centerline{\fbox{\parbox{\columnwidth}{\rule{0cm}{20ex}}}}\fi
\caption{\label{fig-WoE}
  Energy dependence of the quasieigenenergies $\omega$ for two sample
  configurations. Instead of using the quasispectrum obtained from
  $\omega_l(E=0)$ ($\bigcirc $) we calculate the real eigenenergies
  according to $\omega(E_k)=0$ ($\Box $). Different line styles
  distinguish different $\omega_l(E)$. We emphasize that the observed
  behavior varies from sample to sample between remarkably linear (a)
  and strongly nonlinear (b).}
\end{figure}

It is instructive to compare our procedure of calculating the
energy levels with an approach adopted in large-scale simulations
within the CC model.\cite{Met98b,KleM97}  This approach is based on the
unitary network operator $U$.\cite{KleM97} For a single RG unit this
operator acts analogously to the matrix in the left-hand side of Eq.\
(\ref{system}). However, within the approach of Refs.
\onlinecite{Met98b} and \onlinecite{KleM97},
 the energy dependence of phases $\Phi_j$ in
the elements of the matrix was neglected (only the random
contributions, $\Phi_{0,j}$, were kept).  Then, instead of the energy
levels, $E_k$, diagonalization of the matrix (\ref{system})
yielded a set of eigenvalues, $\exp(i\omega_k)$. The numbers
$\omega_k$ were named {\em quasienergies}, and it is the statistics of
these quasienergies that was studied in Ref.\ \onlinecite{KleM97}.
Comparison of the two procedures for a single RG unit is illustrated
in Fig.\ \ref{fig-WoE}.
Fig.\ \ref{fig-WoE}
 shows the dependence of the $4$ quasienergies $\omega_k$ on
the energy $E$ calculated for two single sample RG units, 
with  $t_i$  chosen
from the critical distribution $P_{\text c}(t)$. The energy
dependence of the phases $\Phi_j$ was chosen from LSD of the GUE
according to Eq.\ (\ref{eq-PhiE}).
It is seen that the dependences $\omega(E)$ range from remarkably
linear and almost parallel (Fig.\ \ref{fig-WoE}a) to strongly
nonlinear (Fig.\ \ref{fig-WoE}b).
%
%
\section{Numerical results}
\label{sec-num}
\subsection{The LSD at the QH transition}
As a first step of the RG procedure for the calculation of the critical
LSD we chose for $P_0(s)$ the distribution corresponding to the GUE
random matrix ensemble, since previous simulations\cite{BatS96,KleM97}
indicated that the LSD at the transition is close to GUE.  According
to $P_0(s)$, we pick $s_j$ and set $\Phi_j$, $j= 1,\ldots,4$ as in
Eq.\ (\ref{eq-PhiE}).  For the transmission coefficients $t_i$,
$i=1,\ldots,5$ we use the fixed-point distribution $P_{\text
  c}(t)$,\cite{comment-1} obtained previously.\cite{CaiRSR01}
 
\begin{figure}
\ifshowfigures%
\centerline{\includegraphics[width=0.95\columnwidth]{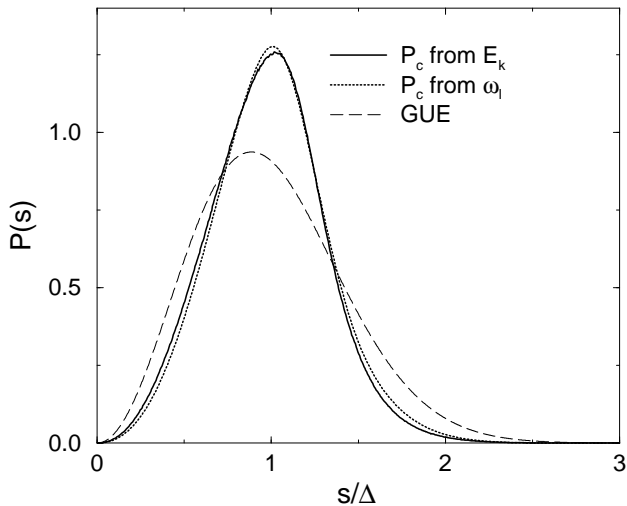}}
\else\centerline{\fbox{\parbox{\columnwidth}{\rule{0cm}{20ex}}}}\fi
\caption{\label{fig-PsFP}
  Critical distributions $P_{\text c}(s)$ obtained from the spectrum of
  $\omega_l(E=0)$ and from the RG approach using the real
  eigenenergies $E_k$ in comparison to the LSD for GUE. As in all
  other graphs $P(s)$ is shown in units of the mean level spacing
  $\Delta$.}
\end{figure}

From the solutions of Eq.\ (\ref{system}) corresponding to
$\omega_j(E_k)=0$ the new LSD $P_1(s')$ is constructed using the
``unfolded'' energy level spacings $s'_m=(E_{m+1}-E_{m})/\Delta$,
where $m=1,2,3$, $E_{k+1}>E_{k}$ and the mean spacing $\Delta=
(E_4-E_1)/3$.  Due to the ``unfolding''\cite{Haa92} with $\Delta$, the
average spacing is set to one for each sample and in each RG-iteration
step we superimpose spacing data of $2\times 10^6$ RG units. The
resulting LSD is discretized in bins with largest width $0.01$.  In
the following iteration step we repeat the procedure using $P_1$ as
initial distribution.  We assume that the iteration process has
converged when the mean-square deviation of distribution $P_n(s)$
deviates by less than $10^{-4}$ from its predecessor $P_{n-1}(s)$.  The RG
iteration process converges rather quickly after only $2-3$ RG steps.
The resulting LSD, $P_{\text c}(s)$, is shown in Fig.\ \ref{fig-PsFP}
together with an LSD for the unitary random matrix ensemble.

Although $P_{\text c}(s)$ exhibits the expected features, namely,
level repulsion for small $s$ and a long tail at large $s$, the
overall shape of $P_{\text c}(s)$ differs noticeably from GUE.  In the
previous large-size lattice simulations \cite{BatS96,KleM97} the
obtained critical LSD was much closer to GUE than $P_{\text c}(s)$ in
Fig.\ \ref{fig-PsFP}. This fact, however, does not reflect on the
accuracy of the RG approach. Indeed, as it was demonstrated recently,
the critical LSD -- although being system size independent ---
nevertheless depends on the geometry of the samples\cite{PotS98} and
on the specific choice of boundary conditions.\cite{BraMP98,SchP98}
Sensitivity to the boundary conditions does not affect the asymptotics
of the critical distribution, but rather manifests itself in the shape
of the ``body'' of the LSD. Recall now that the boundary conditions
which we have imposed to calculate the energy levels (dashed lines in
Fig.\ \ref{fig-RGstruct}) are {\em non-periodic}.

\begin{figure}
\ifshowfigures%
\centerline{\includegraphics[width=0.95\columnwidth]{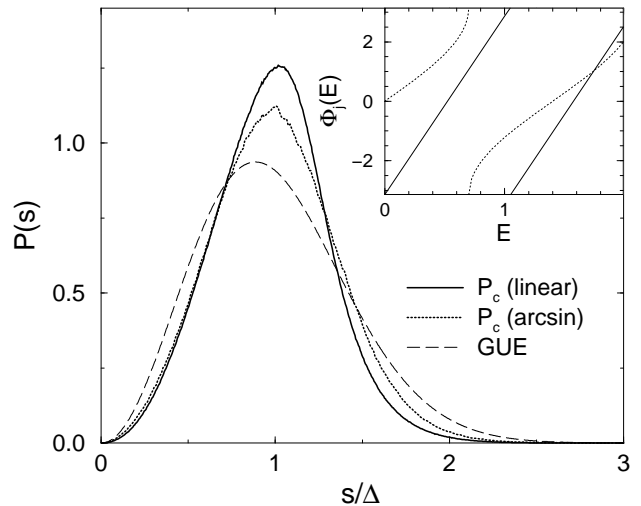}}
\else\centerline{\fbox{\parbox{\columnwidth}{\rule{0cm}{20ex}}}}\fi
\caption{\label{fig-PsFParcsin}
  Critical distributions $P_{\text c}(s)$ for a linear and an $\arcsin$
  energy dependence of the phases $\Phi_j$. The form of $P_{\text
    c}(s)$ is clearly influenced by the actual choice of $\Phi_j(E)$.
   Hence the balk of the distribution is non-universal.  The
  inset illustrates examples of the two different functions
  $\Phi_j(E)$ as in Eqs.\ (\ref{eq-PhiE}) and (\ref{eq-Phi2E}).}
\end{figure}

There is another possibility to assess the critical LSD, namely by
iterating the distribution of {\em quasienergies}.  In Fig.\ 
\ref{fig-PsFP} we show the result of this procedure.  It appears that
the resulting distribution is almost {\em identical} to $P_{\text
  c}(s)$.  This observation is highly non-trivial, since, as follows
from Fig.\ \ref{fig-WoE}, there is no simple relation between the
energies and quasienergies. Moreover, if instead of the linear
$E$-dependence of $\Phi_j$, we choose another functional form, say,
\begin{equation}
\label{eq-Phi2E}
\Phi_j(E)=\Phi_{0,j}+2\arcsin\left(\frac{E}{s_j}-2p\right),
\end{equation}
where the integer $p$ insures that $\left|\frac{E}{s_j}-2p\right|\le
1$, then, the RG procedure would yield an LSD which is markedly
different (within the ``body'') from $P_{\text c}(s)$. This is
illustrated in Fig.\ \ref{fig-PsFParcsin}.
 
Both procedures, using quasienergies instead of real energies [as in
Ref.\ \onlinecite{KleM97}], and linearization of the energy dependence
of phases [as in Eq.\ (\ref{eq-PhiE})] are not rigorous. Linearization
is dictated by the RG concept.  The coincidence of the results
of the two procedures indicates that the concept of quasienergies,
namely, that they obey the same statistics as real energies, is
equivalent to the RG.

\subsection{Small and large $s$ behavior}

\begin{figure}
\ifshowfigures%
\centerline{\includegraphics[width=0.95\columnwidth]{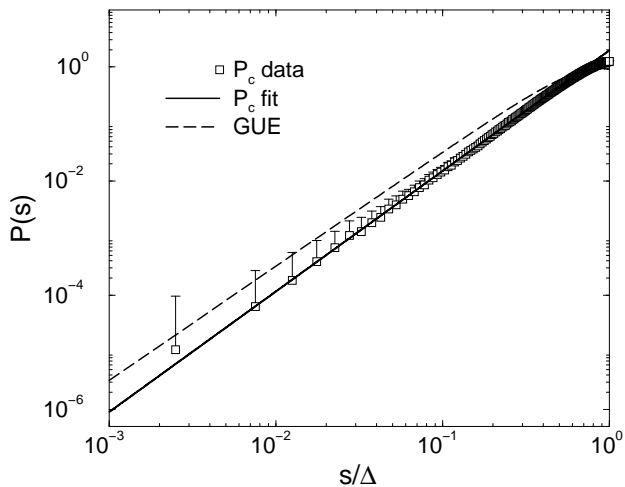}}
\else\centerline{\fbox{\parbox{\columnwidth}{\rule{0cm}{20ex}}}}\fi
\caption{\label{fig-PsFPsmall} 
  Critical $P_{\text c}(s)$ for small $s$ in agreement with the 
predicted
  $s^2$ behavior. Due to the log-log plot errors are shown in the
  upper direction only.  }
\end{figure}

As it was mentioned above, the general shape of the critical LSD is
not universal. However, the small $s$ behavior of $P_{\text c}(s)$
must be the same as for the unitary random matrix ensemble, namely
$P_{\text c}(s) \propto s^2$. This is because delocalization at the
quantum Hall transition implies the level
repulsion.\cite{ShkSSL93,FyoM97} Earlier large-scale simulations of
the critical LSD
\cite{KawOSO96,BatSZK96,Met98b,FeiAB95,OhtO95,BatS96,MetV98,BatSK98,Met99,OnoOK96,KleM97}
satisfy this general requirement.  The same holds also for our result,
as can be seen in Fig.\ \ref{fig-PsFPsmall}. The given error bars of
our numerical data are standard deviations computed from a statistical
average of $100$ FP distributions each obtained for different random
sets of $t_i$'s and $\Phi_j$'s within the RG unit. In general, within
the RG approach, the $s^2$-asymptotics of $P(s)$ is most natural.
This is because the levels are found from diagonalization of the
$4\times 4$ unitary matrix with absolute values of elements widely
distributed between $0$ and $1$.

\begin{figure}
\ifshowfigures%
\centerline{\includegraphics[width=0.95\columnwidth]{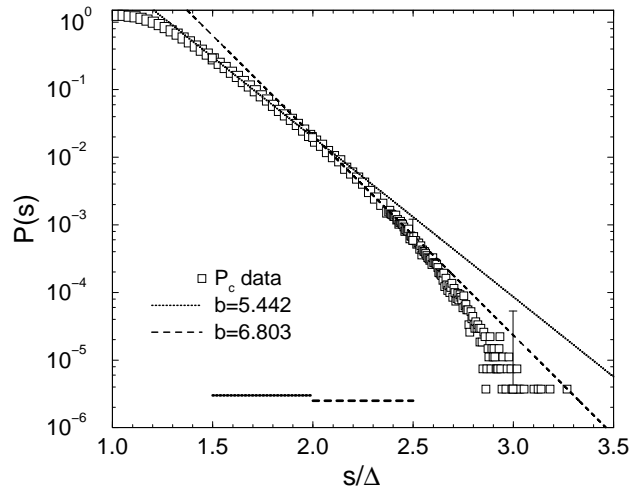}}
\else\centerline{\fbox{\parbox{\columnwidth}{\rule{0cm}{20ex}}}}\fi
 \caption{\label{fig-PsFPlarge} 
   The large $s$ tail of $P_{\text c}(s)$ compared with fits
   according to the predictions of Ref.\ \protect\onlinecite{ShkSSL93}
   (lines). The interval used for fitting is indicated by the
   bars close to the lower axis. For clarity errors are shown in upper
   direction and for $s/\Delta=1.5,2.0,2.5,3.0$ only. 
For $s/\Delta<2.4$, only
   every $5$th data point is drawn by a symbol.}
 \end{figure}

The right form of the large-$s$ tail of $P(s)$ is Poissonian,
$P_{\text c}(s)\propto \exp(-bs)$.\cite{ShkSSL93} For the Anderson
model in three dimensions, unambiguous confirmation of this prediction
in numerical simulations became possible only when very high numerical
accuracy had been achieved.\cite{BatSZK96,ZhaK97} This is because
$P_{\text c}(s)$ assumes the Poissonian asymtotics only at large
enough $s \gtrsim 3\Delta$.  For the quantum Hall transition, a linear
behavior of $\ln P_{\text c}(s)$ with a slope corresponding to the
value $b\approx 4.1$ has been found in Ref.\ \onlinecite{BatS96} from
the analysis of the interval $2< s/\Delta <4$.  Our data, as shown in
Fig.\ \ref{fig-PsFPlarge}, has a high accuracy only for $s/\Delta
\lesssim 2.5$. For such $s$, the distribution $P_{\text c}(s)$ does
not yet reach its large-$s$ tail.  Thus, the value of parameter $b$
extracted from this limited interval is somewhat ambiguous. Namely, we
obtain $b=5.442$ for $s/\Delta\in[1.5,2.0]$ and $b=6.803$ for
$s/\Delta\in[2.0,2.5]$.

Summarizing, the accuracy of the RG approach, applied to the level
statistics, is insufficient to discern the only non-trivial feature of
the critical LSD, i.e., the universal Poissonian asymtotics.  However,
the scaling analysis of LSD clearly reveals the universal features of
the quantum Hall transition as we demonstrate in the next Section.

\section{Scaling results for the LSD}
\label{sec-FSS}

\subsection{Finite-size scaling at the QH transition}

The critical exponent, $\nu$, of the quantum Hall transition governs
the divergence of the correlation length $\xi_\infty$ as a function of
the arbitrary control parameter $z_0$, i.e., 
\begin{equation}
\label{eq-diverge}
\xi_\infty(z_0) \propto |z_0-z_{\text c}|^{-\nu},
\end{equation}
where $z_{\text c}$ is the critical value. The values of $\nu$
calculated using different numerical methods, e.g.,
$\nu=2.35\pm0.03$,\cite{Huc92} $2.4\pm0.2$,\cite{LeeWK93}
$2.5\pm0.5$\cite{ChaC88} agree with each other.  The RG approach for
the conductance distribution also yields a rather accurate value $\nu
=2.39\pm0.01$.\cite{CaiRSR01} In Sec. II we have introduced a
complimentary RG approach to the distribution of the energy levels at
the transition.  It can be expected on general grounds, that the LSD
obtained from the RG approach obeys scaling at small enough
$z_0-z_{\text c}$. However, it is by now means obvious, whether the
value of $\nu$ extracted from different variants of the RG approach
are consistent.

In order to extract $\nu$ from the LSD we employ the
one-parameter-scaling analysis. This analysis
is based on the rescaling of a quantity $\alpha(N; \{z_i\})$ ---
depending on (external) system parameters $\{z_i\}$ and the system
size $N$ --- onto a single curve by using a scaling function $f$
\begin{equation}
\label{eq-scal}
\alpha\left(N;\{z_i\}\right)=f\left(\frac{N}
{\xi_\infty(\{z_i\})}\right) .
\end{equation}  
Since Eq.\ (\ref{eq-diverge}), as indicated by ``$\infty$'', holds
only in the limit of infinite system size, we now use the scaling
assumption to extrapolate $f$ to $N\rightarrow\infty$ from the
finite-size results of the computations.  Once $f$ and
$\xi_\infty$ are known, the value of $\nu$ can be then inferred.

In the original formulation of the RG approach\cite{GalR97} it was
demonstrated that there is a natural parametrization of the
transmission coefficients $t$, i.e., $t=(e^z+1)^{-1/2}$.  For such a
parametrization, $z$ can be identified with a dimensionless electron
energy. The quantum Hall transition occurs at $z=0$, which corresponds
to the center of the Landau band. The universal conductance
distribution at the transition, $P_{\text c}(G)$, corresponds to the
distribution $Q_{\text c}(z)=P_{\text c}\left[(e^z+1)^{-1}\right] /
{4}\cosh^{2}(z/2)$ of parameter $z$, which is symmetric with respect
to $z=0$ and has a shape close to a gaussian.\cite{CaiRSR01} The RG
procedure for the conductance distribution converges and yields $Q_{\text
  c}(z)$ only if the initial distribution is an even function of $z$.
This suggests to choose as a control parameter in Eq.\ 
(\ref{eq-scal}), $z_0$, the position of the maximum of the function
$Q(z)$.  Then the meaning of $z_0$ is the electron energy measured
from the center of Landau band. The fact that the quantum Hall
transition is infinitely sharp implies that for any $z_0\ne 0$, the RG
procedure drives the initial distribution $Q(z-z_0)$ towards an
insulator, either with complete transmission of the network nodes (for
$z_0>0$) or with complete reflection of the nodes (for $z_0<0$).

\subsection{Scaling for $\alpha_{\text P}$ and $\alpha_{\text I}$}

\begin{figure}
\ifshowfigures%
\centerline{\includegraphics[width=0.95\columnwidth]{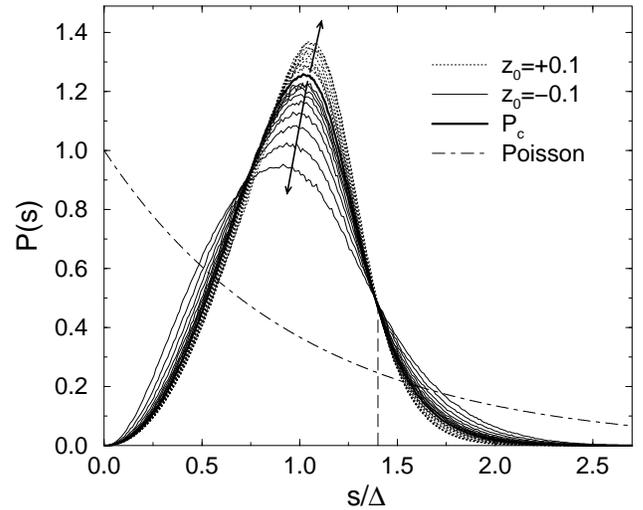}}
\else\centerline{\fbox{\parbox{\columnwidth}{\rule{0cm}{20ex}}}}\fi
\caption{\label{fig-LSDshift}
  RG of the LSD used for the computation of $\nu$. The dotted lines
  corresponds to the first $9$ RG iterations with an initial
  distribution $P_0$ shifted to complete transmission ($z_0=0.1$) while
  the long-dashed lines represent results for a shift toward
  complete reflection ($z_0=-0.1$). Within the RG procedure
 the LSD moves
  away from the FP as indicated by the arrows. At $s/\Delta\approx
  1.4$ the curves cross at the same point -- a feature we exploit
  when deriving a scaling quantity from the LSD.  }
\end{figure}

In principle, we are free to choose for the finite-size analysis any
characteristic quantity $\alpha(N;z_0)$ constructed from the LSD which
has a systematic dependence on system size $N$ for $z_0\ne0$ while
being constant at the transition $z_0=0$.  Because of the large number
of possible
choices\cite{ShkSSL93,HofS94b,ZhaK97,BatS96,ZhaK95b,ZhaK95c} we
restrict ourselves to two quantities which are obtained by integration
of the LSD and have already been successfully used in Refs.\ 
\onlinecite{HofS94b} and \onlinecite{HofS93}, namely
\begin{equation}
  \alpha_{\text P} = \int^{s_0}_0 P(s) ds
\end{equation}
and second
\begin{equation}
  \alpha_{\text I} = \frac{1}{s_0}\int^{s_0}_0 I(s)ds ,
\end{equation}
with $I(s)= \int^{s}_0 P(s') ds'$. The integration limit is chosen as
$s_0= 1.4$ which approximates the common crossing point\cite{HofS94b}
of all LSD curves as can be seen in Fig.\ \ref{fig-LSDshift}.
Thus $P(s_0)$ is independent of the distance $|z_0-z_{\text c}|$ to
the critical point and the system size $N$. We note that $N$ is
directly related to the RG step $n$ by $N=2^n$.
The double integration in $\alpha_{\text I}$ is numerically
advantageous since fluctuations in $P(s)$ are smoothed.
We now apply the finite-size-scaling approach from Eq.\ 
(\ref{eq-scal})
\begin{equation}
  \alpha_{\text{I,P}}(N,z_0)= f\left(\frac{N}{\xi_\infty(z_0)}\right) .
\end{equation}
Since $\alpha_{\text{I,P}}(N,z_0)$ is analytical for finite $N$, one
can expand the scaling function $f$ at the critical point. The first
order approximation yields
\begin{equation}
  \alpha(N,z_0)\approx \alpha(N,z_{\text c})+ 
a |z_0-z_{\text c}| N^{1/\nu}
\end{equation}
where $a$ is a dimensionless coefficient.  For our calculation we use
a higher order expansion proposed by Slevin and Ohtsuki.\cite{SleO99a}
In Ref.\ \onlinecite{SleO99a} the function $f$ is expanded twice,
first, in terms of the Chebyshev polynomials of order ${\cal O}_\nu$
and, second, in Taylor series with terms $|z_0-z_c|$ in the power
${\cal O}_z$.  This procedure allows to describe the deviations from
linearity in $|z_0-z_c|$ at the transition.  In addition, in Ref.\ 
\onlinecite{SleO99a} the contributions from an irrelevant scaling
variable which leads to a shift of the transition for small system
sizes was taken into account.  In our case, in contrast to the
Anderson model of localization, the transition point $z_0=0$ is known.
\begin{figure}
\ifshowfigures
\centerline{\includegraphics[width=0.85\columnwidth]{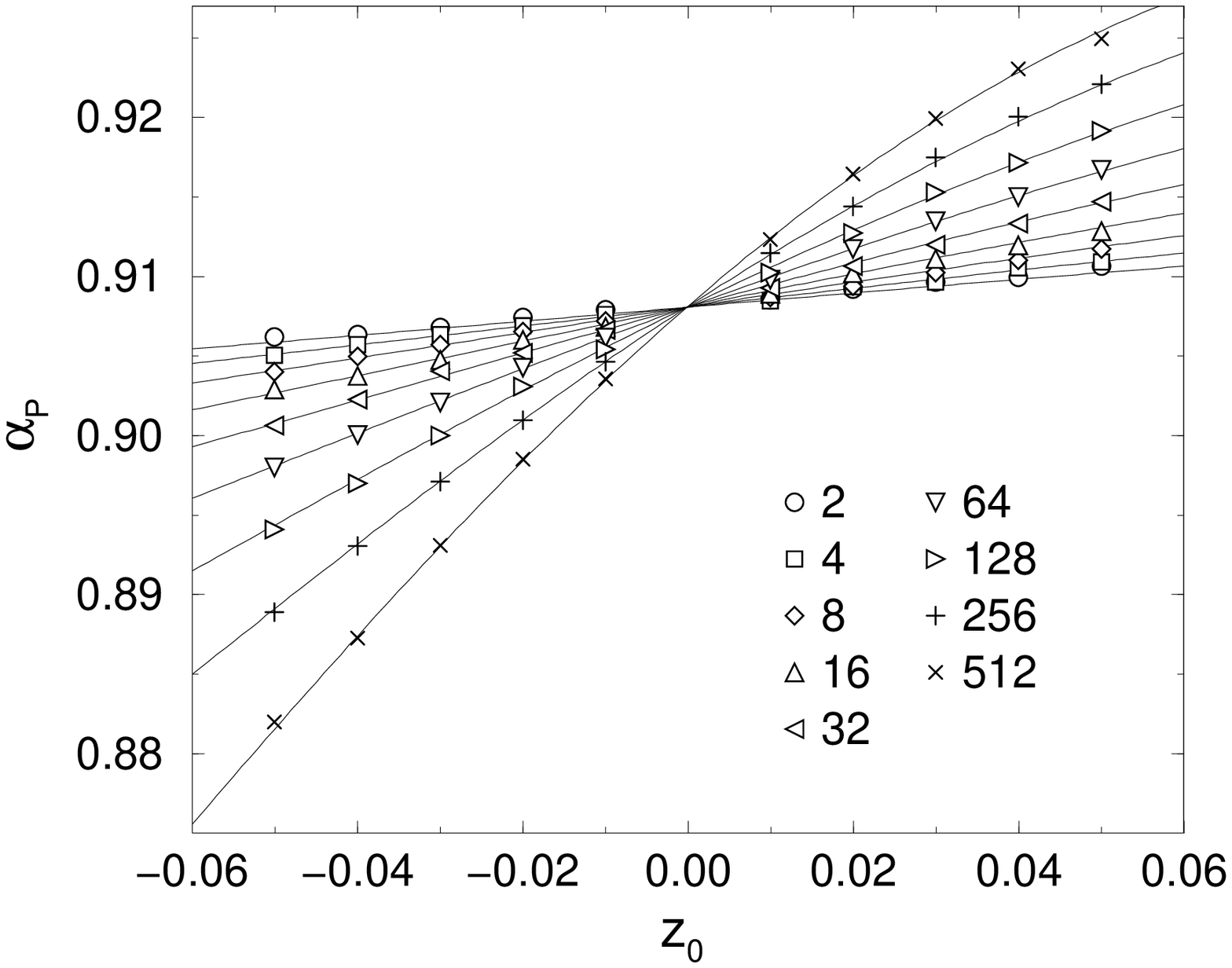}}
\centerline{\includegraphics[width=0.85\columnwidth]{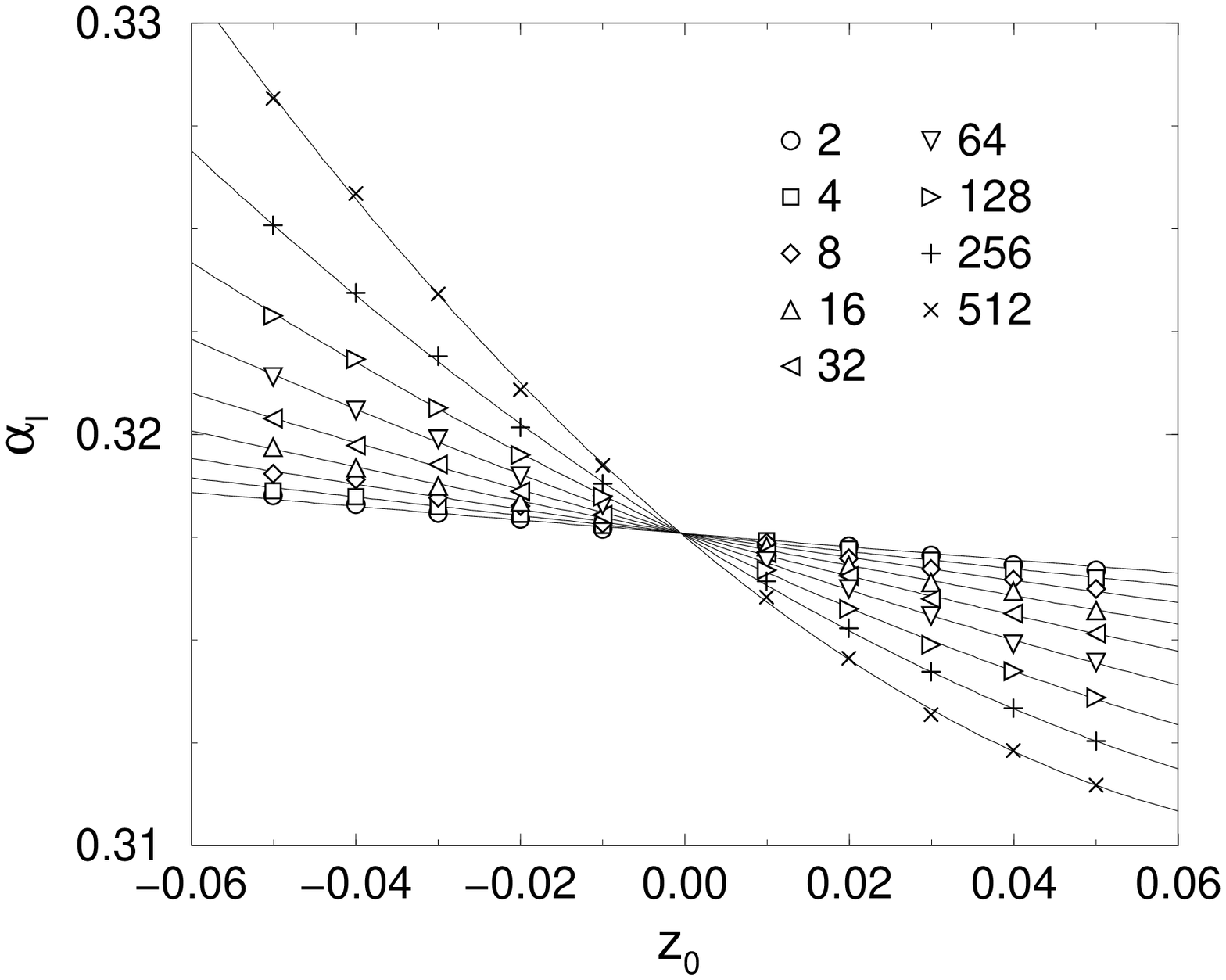}}
\else\centerline{\fbox{\parbox{\columnwidth}{\rule{0cm}{20ex}}}}\fi
\caption{\label{fig-AlphaVsZ0}
  Behavior of $\alpha_{\text I}$ and $\alpha_{\text P} $ at the QH
  transition as results of the RG of the LSD. Data are shown for RG
  iterations $n=1,\ldots,9$ corresponding to effective system sizes
  $N=2^n=2,\ldots,512$. Full lines indicate the functional dependence
  according to FSS using the $\chi^2$ minimization with ${\cal
    O}_{\nu}=2$ and ${\cal O}_{z}=3$.  }
\end{figure}
Therefore, we can neglect the influence of irrelevant variables. In
order to obtain the functional form of $f$, the fitting parameters,
including $\nu$, are evaluated by a nonlinear least-square ($\chi^2$)
minimization. In Fig.\ \ref{fig-AlphaVsZ0} we show the resulting fit
for $\alpha_{\text P}$ and $\alpha_{\text I}$ at the transition.
%
\begin{figure}
\ifshowfigures
\centerline{\includegraphics[width=0.85\columnwidth]{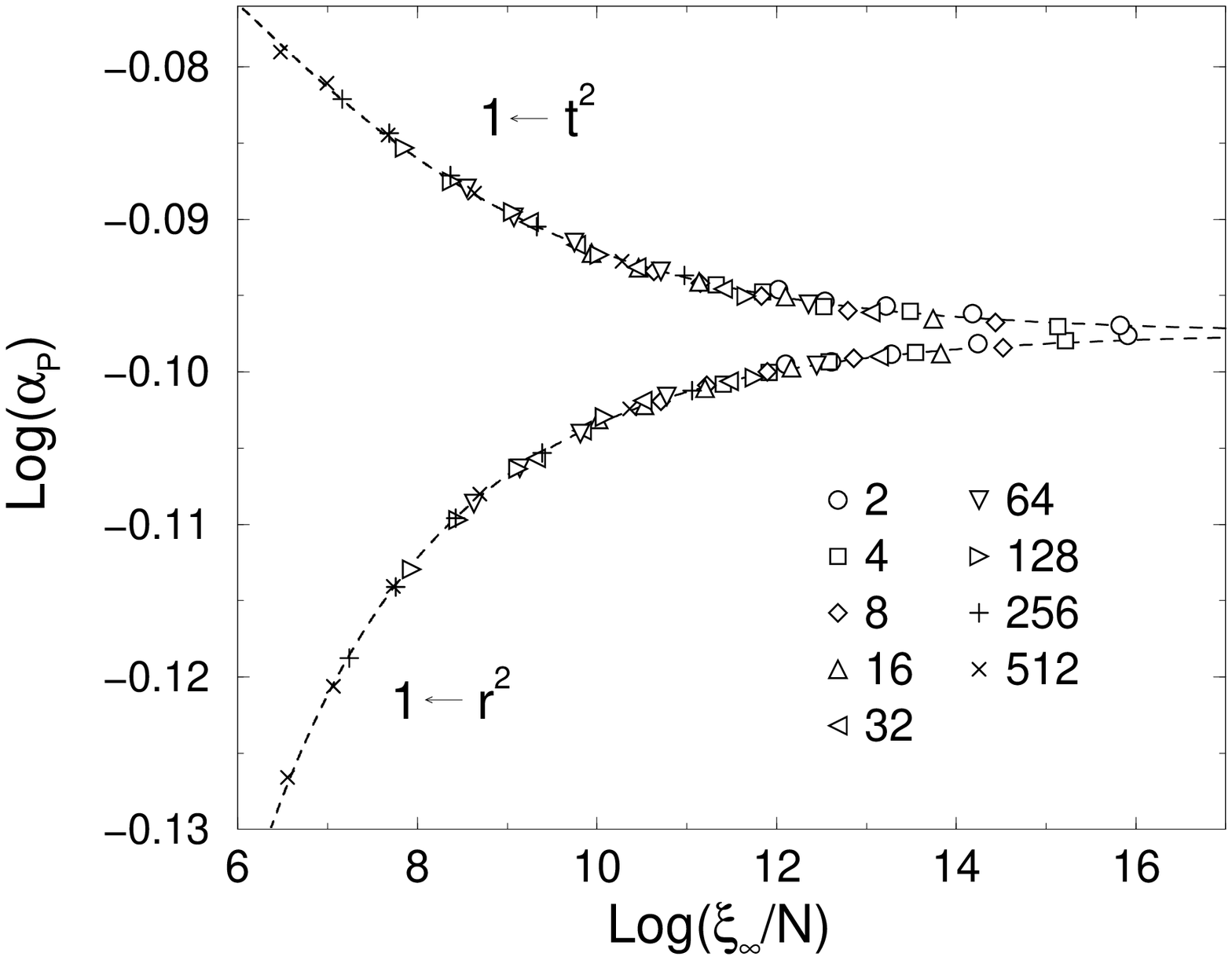}}
\centerline{\includegraphics[width=0.85\columnwidth]{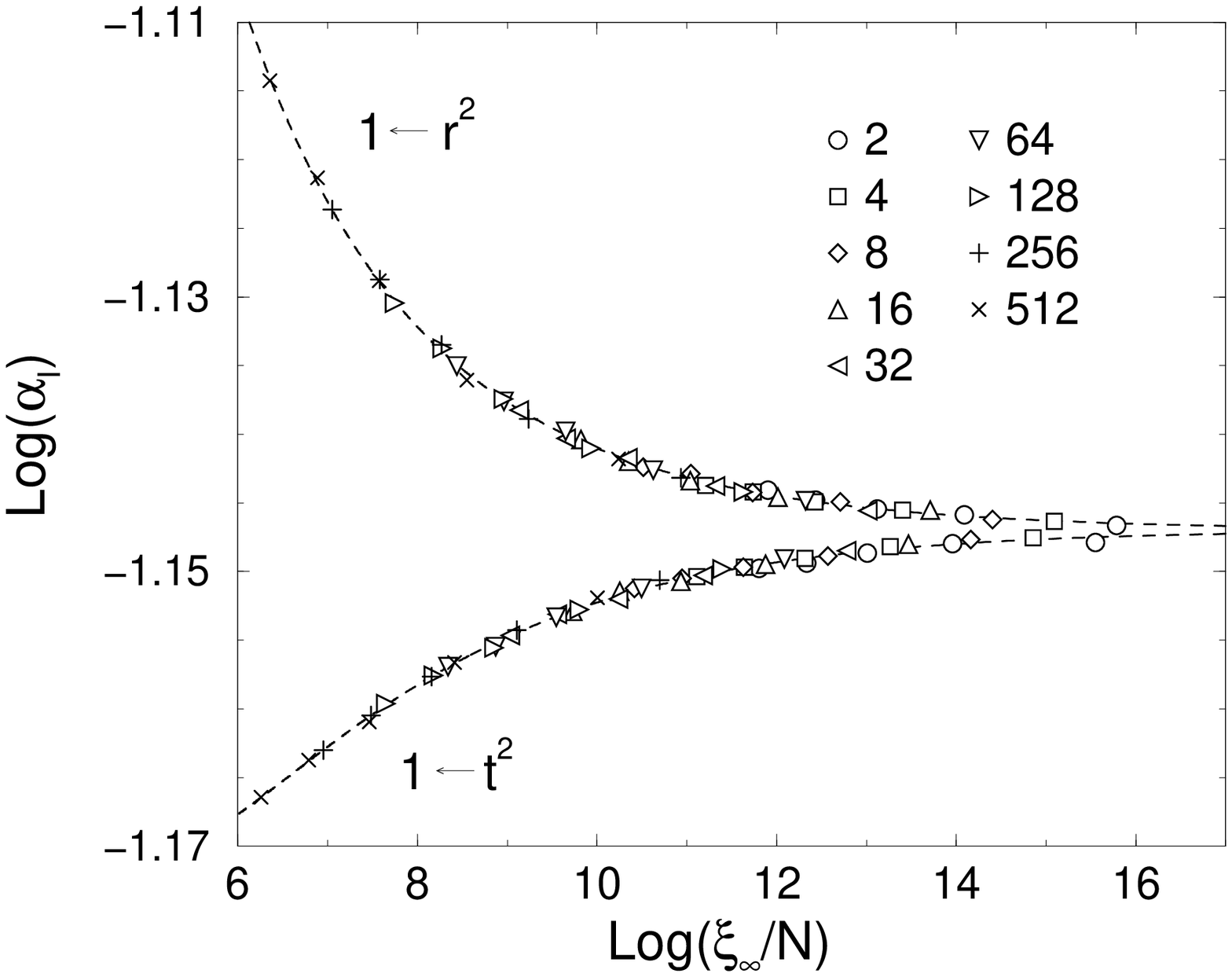}}
\else\centerline{\fbox{\parbox{\columnwidth}{\rule{0cm}{20ex}}}}\fi
\caption{\label{fig-AlphaFSS} 
  Finite size scaling curves resulting from the $\chi^2$ fit of our
  data shown in Fig.\ \protect\ref{fig-AlphaVsZ0}.  Different symbols
  correspond to different effective system sizes $N=2^n$.  The data
  points collapse onto a single curve indicating the validity of the
  scaling approach.}
\end{figure}

The fits are chosen in such a way that the total number of parameters
is kept at a minimal value, while the fit agrees well with the
numerical data.\cite{NdaRS02} The corresponding scaling curves are
displayed in Fig.\ \ref{fig-AlphaFSS}. In the plots the two branches
corresponding to complete reflection ($z_0<0$) and complete
transmission ($z_0>0$) can be clearly distinguished.  In order to
estimate the error of fitting procedure we compare the results for
$\nu$ obtained by different orders ${\cal O}_\nu$ and ${\cal O}_z$ of
the expansion, system sizes $N$, and regions around the transition.  A
part of our over $100$ fit results together with the standard
deviation of the fit are given in Table \ref{tab-nucomp}.
\begin{table}
\caption{ \label{tab-nucomp} Part of fit results for $\nu$ obtained
  from $\alpha_{\text I}$ and $\alpha_{\text P}$ for different system
  sizes $N$, intervals around the transition, orders ${\cal O}_\nu$
  and ${\cal O}_z$ of the fitting procedure.}
\begin{tabular}{ccccc}
  $N$&$[z_{0\text{min}},z_{0\text{max}}]$&${\cal O}_\nu$ & ${\cal O}_z$ & 
$\nu$\\ 
  \multicolumn{5}{c}{$\alpha_{\text P}$}\\\hline
  $2-512$&$[9.93,10.07]$&$3$&$2$&$2.336\pm0.010$\\ 
  $2-256$&$[9.93,10.07]$&$2$&$3$&$2.412\pm0.013$\\ 
  $4-512$&$[9.95,10.05]$&$3$&$1$&$2.325\pm0.014$\\ 
  $2-512$&$[9.95,10.05]$&$2$&$1$&$2.402\pm0.014$\\ 
  $2-256$&$[9.95,10.05]$&$2$&$2$&$2.360\pm0.016$\\ 
  $16-512$&$[9.95,10.05]$&$2$&$3$&$2.385\pm0.018$\\ 
  $2-128$&$[9.93,10.07]$&$1$&$3$&$2.384\pm0.019$\\ 
  $4-512$&$[9.93,10.07]$&$2$&$1$&$2.471\pm0.019$\\ 
  \multicolumn{5}{c}{$\alpha_{\text I}$}\\\hline
  $2-512$&$[9.93,10.07]$&$2$&$2$&$2.383\pm0.010$\\ 
  $2-512$&$[9.93,10.07]$&$2$&$3$&$2.388\pm0.010$\\ 
  $2-512$&$[9.93,10.07]$&$3$&$1$&$2.346\pm0.012$\\ 
  $8-512$&$[9.93,10.07]$&$2$&$3$&$2.376\pm0.012$\\ 
  $2-512$&$[9.95,10.05]$&$2$&$3$&$2.368\pm0.014$\\ 
  $2-128$&$[9.93,10.07]$&$2$&$3$&$2.377\pm0.016$\\ 
  $16-512$&$[9.95,10.05]$&$2$&$1$&$2.367\pm0.016$\\ 
  $2-256$&$[9.93,10.07]$&$3$&$3$&$2.372\pm0.018$\\ 
\end{tabular}
\end{table}
The value of $\nu$ is calculated as the average of all individual fits
where the resulting error of $\nu$ was smaller than $0.02$. The error
is then determined as the standard deviation of the contributing
values. By this method we assure that our result is not influenced by
local minima of the nonlinear fit.  So we consider $\nu=2.37\pm0.02$
as a reliable value for the exponent of the localization length at the
QH transition obtained from the RG approach to LSD.  This is in
excellent agreement with $\nu=2.35\pm0.03$ (Ref.\ \onlinecite{Huc92}),
$2.4\pm0.2$ (Ref.\ \onlinecite{LeeWK93}), $2.5\pm0.5$ (Ref.\ 
\onlinecite{ChaC88}), and $2.39\pm0.01$ (Ref.\ \onlinecite{CaiRSR01})
calculated previously.  In addition to $\alpha_{\text P}$ and
$\alpha_{\text I}$, we tested also a parameter-free scaling quantity
$\int_0^\infty s^2 P(s) ds$,\cite{ZhaK95b} where the whole
distribution $P(s)$ is taken into account. Here, due to the influence
of the large $s$-tail a less reliable value $\nu=2.33\pm0.05$ was
obtained.

\subsection{Test of consistency}

\begin{figure}
\ifshowfigures%
\centerline{\includegraphics[width=0.95\columnwidth]{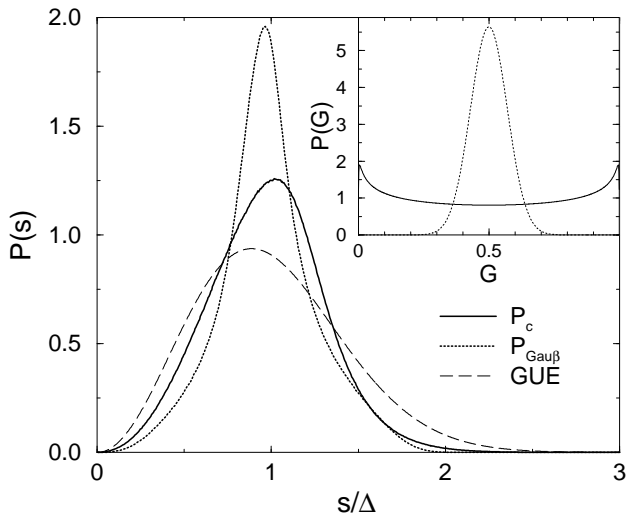}}
\else\centerline{\fbox{\parbox{\columnwidth}{\rule{0cm}{20ex}}}}\fi
  \caption{\label{fig-PsFPcomp} 
    Comparison of the LSD $P_{\text c}(s)$ and $P_{\text{Gau{\ss}}}(s)$
    obtained from the corresponding conductance distributions shown in
    the inset.}
\end{figure}

\begin{figure}
\ifshowfigures%
\centerline{\includegraphics[width=0.95\columnwidth]{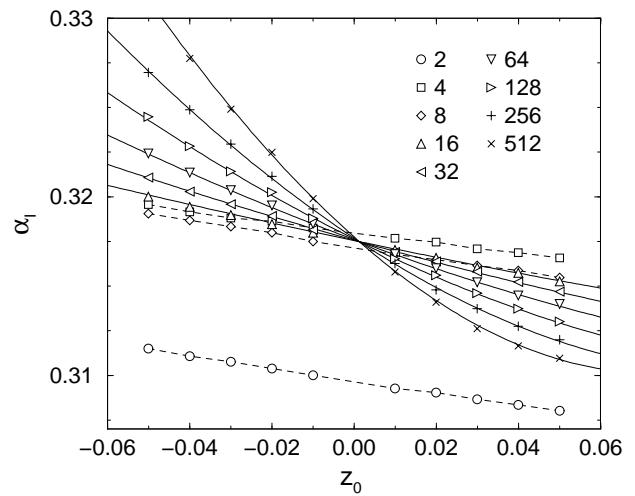}}
\else\centerline{\fbox{\parbox{\columnwidth}{\rule{0cm}{20ex}}}}\fi
  \caption{\label{fig-AlphaVsZ0gauss}
    Behavior of $\alpha_{\text I}$ computed for initial distributions
    $P_{\text{Gau{\ss}}}$ different from the critical distributions,
    as shown in Fig.\ \protect\ref{fig-PsFPcomp}. Data are plotted for
    RG iterations $n=1,\ldots,9$ corresponding to effective system
    sizes $N=2^n=2,\ldots,512$. Curves for small $n$ do not cross at
    the common point $z_0=0$. Full lines indicate the functional
    dependence according to FSS using the $\chi^2$ minimization with
    ${\cal O}_{\nu}=2$ and ${\cal O}_{z}=2$.}
\end{figure}

Finally we address the question, how the actual form of the
distribution $Q(z)$ affects the results for LSD and the scaling
analysis.  Recall that in the above calculations we have used at each
step of the RG procedure the distribution $Q(z)$ derived from the
critical conductance distribution, $P_{\text c}(G)$.  The function
$P_{\text c}(G)$ is shown in Fig.\ \ref{fig-PsFPcomp} (inset) with a
full line.  In order to understand the importance of the fact that
$P_{\text c}(G)$ is almost flat, we have repeated our calculations
choosing for $P(G)$ a relatively narrow gaussian distribution
$P(G)\equiv P_{\text{Gau{\ss}}}(G)$ at each RG step.  This
distribution is shown with a dashed line in Fig.\ \ref{fig-PsFPcomp}
(inset). The obtained LSD is presented in Fig.\ \ref{fig-PsFPcomp}.
Obviously, it agrees much worse with GUE, which can be considered as a
reference point, than the LSD computed using the true $P_{\text
  c}(G)$.  Our data for $\alpha_{\text I}$ calculated for
$P(G)=P_{\text{Gau{\ss}}}(G)$ is plotted in Fig.\ 
\ref{fig-AlphaVsZ0gauss}.
%
The curves for small system sizes $N$ exhibit strong deviations, i.e.,
there is initially no common crossing point, while for large $N$ a
behavior similar to Fig.\ \ref{fig-AlphaVsZ0} is observed.  Therefore,
small $N$ data are neglected in the scaling analysis.  The $\chi^2$
fits for $\alpha_{\text I}$ and $\alpha_{\text P}$ are carried out
using $z_0\in[-0.05,0.05]$ and $N=16-512$.  They yield the values
$\nu_{\text I}=2.43\pm0.02$ and $\nu_{\text P}=2.46\pm0.03$, which are
also less accurate than $\nu$ calculated with the critical $P_{\text
  c}(G)$. Overall, Figs. 10 and 11 illustrate the consistency of the
RG approaches for the conduction distribution and for the level
statistics, in the sense, that the best fixed point distribution of
the level spacings corresponds to the fixed point of the conductance
distribution.


\section{Conclusion}
\label{sec-sum}

Network models introduced in Ref.\ \onlinecite{Sha82} turned out to be
a powerful tool to study the Anderson localization.  Without magnetic
field, the propagation of electron waves along each link of the
network is allowed in both directions.  In two dimensions the
transmission coefficient of the network is zero for all parameters of
the scattering matrix at the nodes,\cite{FreJM99} illustrating
complete localization of electronic states. On the other hand, the
two-channel network model with inter-channel mixing, that models
spin-orbit interaction, exhibits a localization-delocalization
transition\cite{MerJH98} that is also in accord with the scaling
theory of localization.\cite{HikLN80} However, the version of the
network model that has been most widely studied, is the chiral
version, i.e., the CC model,\cite{ChaC88} describing the electron
motion in a disordered system in a strong magnetic field limit. Within
the CC model, the scattering matrix at the node is parametrized by a
single number, e.g., the transmission coefficient $t$.  On the
qualitative level, the CC model yields a transparent explanation why
delocalization occurs only at a single energy, for which $t^2=1/2$.
On the quantitative level, in addition to the exponent, $\nu$, more
delicate characteristics of the critical wave functions were extracted
from the numerical analysis of the CC model.\cite{HucK99,KleZ01}

The fact that the RG approach, within which the correlations between
different scales are neglected, describes the results of the
large-scale simulations of the CC model so accurately, indicates that
only a few spatial correlations within each scale are responsible for
the critical characteristics of the quantum Hall transition. More
precisely, the structure of the eigenstates of a macroscopic sample at
the transition can be predicted from the analysis of a single RG unit
consisting of only five nodes. Earlier we have demonstrated this fact
for the conductance distribution.\cite{CaiRSR01} In the present paper
this statement is reinforced by the study of the level statistics at
the transition, which is a complimentary (to the conductance
distribution) characteristics of the localization.
   
\acknowledgments We thank R.\ Klesse, L.\ Schweitzer and I.\ 
Zharekeshev for stimulating discussions.  This work was supported by
the NSF-DAAD collaborative research Grant No.\ INT-0003710.  P.C.\ and
R.A.R.\ also gratefully acknowledge the support of DFG within the
Schwer\-punkt\-pro\-gramm ``Quanten-Hall-Systeme'' and the SFB~393.



\begin{thebibliography}{10}
\bibitem{comment-1} In a ``full'' RG approach the distributions $P(t)$
  and $P(s)$ should be iterated simultaneously, which is possible but
  has proven to be numerically less stable. Therefore $P_{\text c}(t)$
  is used in all RG steps without affecting the results.
%
  We also note that the chosen form of $P_0(s)$ does not change
  $P_{\text c}(s)$ but might alter the speed of convergence.
\end{thebibliography}
\end{document}